# Taming Randomness in Random Lasers: Programmable Disorder for Active Control of Random Lasing via Electric-Field-Directed Assembly of Nanowires

**Authors**
*Jinkai Yang,[1,2] Kumudu N. Ranasinghe,[3] Lei Kang,[1,*] Jennifer R. Decker,[3] Cheng-Yu Wang,[1,2] Douglas H. Werner,[1] Christine D. Keating[3,*] and Zhiwen Liu[1,2,*]*

**Affiliations**
1. School of Electrical Engineering and Computer Science, The Pennsylvania State University, University Park, Pennsylvania 16802, United States.
2. Materials Research Institute, The Pennsylvania State University, University Park, Pennsylvania 16802, United States.
3. Department of Chemistry, The Pennsylvania State University, University Park, Pennsylvania 16802, United States.

**Email Address**

Jinkai Yang: jxy208@psu.edu
Kumudu N. Ranasinghe: knr5313@psu.edu
Lei Kang: lzk12@psu.edu
Jennifer R. Decker: jrdecker01@ysu.edu
Cheng-Yu Wang: cuw70@psu.edu
Douglas H. Werner: dhw@psu.edu
Christine D. Keating: keating@chem.psu.edu
Zhiwen Liu: zzl1@psu.edu

**\* Corresponding authors**
Lei Kang: lzk12@psu.edu
Christine D. Keating: keating@chem.psu.edu
Zhiwen Liu: zzl1@psu.edu

**Abstract**

Random lasing exploits multiple scattering to provide optical feedback without conventional resonant cavities, enabling simplified architectures that are readily integrated into compact photonic platforms such as wearable sensors and lab-on-chip devices. However, the same disorder that enables cavity-free lasing also makes it challenging to control and tune the emission properties. Here, an electrically reconfigurable random-lasing platform based on dielectrophoretic assembly of chaining silver nanowires suspended in a dye gain medium is reported. An applied electric field across patterned quadrupole electrodes induces nanowire chaining and programmable alignment, enabling real-time reconfiguration of the disorder landscape. Based on the electrically driven disorder state transitions, tunable random-lasing characteristics, including reduced lasing thresholds, modulation of emission intensity, and control of the polarization state have been demonstrated. Simulations further indicate that chaining enhances scattering relative to absorption, providing more efficient radiative feedback, and the orientation of the nanowire network governs the polarization dependence of the system. These results establish a route to actively modulate random lasing through controllable disorder and point toward adaptive, reconfigurable photonic light sources and sensing systems.

## 1. Introduction

In conventional lasers, optical feedback is provided by a fixed resonator, typically formed by mirrors, which traps photons within the gain region and promotes net amplification. Random lasers depart from this paradigm.[1–4] Optical feedback instead emerges from multiple scattering by disordered scatterers embedded in a gain medium. In this sense, the disordered scatterers collectively play the role of the cavity; each scattering event redirects photons and light can remain effectively confined within the gain region for an extended time, allowing lasing without external mirrors.[5–9] The absence of a fixed resonator relaxes alignment and fabrication constraints, enabling low-cost and compact implementations.[10–12] Random lasers naturally operate in disordered media, where multiple scattering both enhances light-matter interaction and makes the optical response highly sensitive to microscopic structural variations.[13,14] This sensitivity, for example, allows possibilities for sensing, particularly in biologically and medically relevant applications.[15–18] In addition, compared with conventional coherent sources, the emission from random lasers can reduce speckle artifacts, motivating studies in speckle-free imaging and related photonic technologies.[19–22] However, the random scattering mechanism that provides optical feedback also brings challenges in controlling random lasers, in part due to the sensitivity of the collective random lasing response to the microstructural configurations of the scatterers. For example, the lasing threshold and the output polarization state often lack tunability.

Control over the random laser can be achieved by manipulating the gain landscape through spatial or temporal shaping of the pump beam or pulse profile within a static random medium, thereby enhancing emission directionality or selecting specific lasing modes.[23–27] Since the random laser feedback mechanism is determined by the disorder configuration that governs light scattering and confinement, an alternative strategy is to engineer the scattering landscape. Mechanical deformation and microstructural shaping were explored to tune the scattering medium, but these methods can suffer from limited reversibility and restricted operational cycling.[28–31]

In previous work, we introduced a dynamically reconfigurable disorder platform to tune random-laser emission characteristics via electric-field-driven assembly of anisotropic dielectric scatterers.[32] Specifically, we used a suspension of titanium dioxide ($TiO_2$) nanowires in a dye solution, where an applied electric field enabled real-time control of nanowire orientation. Because the scattering response of an anisotropic particle depends on its orientation, light incident parallel to a nanowire's long axis can experience a different scattering cross-section than light incident perpendicular to the axis.[33,34] Consequently, field-controlled rotation of the nanowires modulates their scattering interactions and reshapes the effective feedback landscape, enabling dynamic tuning of random-lasing characteristics. In dielectrophoretic assembly, the achievable reconfiguration is strongly influenced by particle polarizability. The dielectric ($TiO_2$) nanowire in our prior work experienced weak dielectrophoretic forces, resulting in only a weak anisotropy and limited structural reconfiguration. Further, dielectric nanowires have limited capability to engineer the near field, which is critical for controlling stimulated emission.

In this study, we use dielectrophoretic actuation to induce dynamic chaining of metallic nanowires, specifically silver (Ag) nanowires, inside a random laser gain medium, leveraging their efficient field response and strong optical scattering.[35–44] The field-driven chaining process does not merely rotate individual scatterers; it actively restructures the nanowire network and thereby reconfigures the disorder topology that governs multiple-scattering pathways.[45–47] The chained nanowires can further selectively enhance the scattering near field produced by an incident wave polarized parallel to the chaining direction.

The chaining and the resultant structural reconfiguration modify local field distribution that underpins random lasing, enabling real-time optical property control across distinct disorder states. As a result, we achieve tunable random-laser characteristics, including a controllable lasing threshold, modulation of emission intensity over orders of magnitude, and electrical control of the polarization state of the random-lasing output.

## 2. Results and Discussion

### 2.1 Electrically reconfigurable random lasing system

**Figure 1** illustrates the central concept of this work: the optical response of a random laser is governed by its internal scattering structure, which can be actively and reversibly reconfigured using an external electric field. The device consists of a dye-doped gain layer containing silver nanowires, sandwiched between two glass slides and integrated with planar electrodes. When an external electric field is applied across the electrodes, Ag nanowires undergo dielectrophoretically induced chaining and alignment, and thereby the internal scattering landscape is reconfigured (See Supporting Information Figure S1). Microscope observations reveal two dominant consequences of this chaining process. First, chained assemblies show increased local nanowire density by nearly a factor of two compared to the unchained state (See Supporting Information, Figure S2). This density increase enhances light–nanowire interactions and the overall scattering strength. Second, chaining introduces pronounced structural anisotropy by converting an initially quasi-isotropic system into an aligned, strongly anisotropic network. In the quasi-isotropic state, nanowires can be considered as randomly oriented in three dimensions. As a result, the incident light interacts with a broad distribution of projected geometries, reducing the coupling efficiency. After chaining and alignment, the nanowires preferentially share a common in-plane direction, leading to stronger interaction with linearly polarized incident light. Reflecting the increased degree of disorder, this transition from isotropy to anisotropy creates an orientation-selective scattering environment and directly alters the distributed optical feedback.

The optical microscopy images exhibiting the observed structural reconfiguration are shown in the top row of Figure 1b. In the absence of an applied field, the nanowires remain randomly oriented, forming an isotropic disordered network. Upon electrical activation, the nanowires assemble into elongated chain networks aligned along either the x or y direction, depending on the field orientation. Two-dimensional autocorrelation analysis (see the inserts in Fig. 1b) provides a quantitative assessment of this transition (See Materials and Methods: Data Collection and Processing), revealing a clear progression from isotropic to strongly anisotropic, directionally ordered scattering networks as electrical assembly proceeds. The bottom row of Figure 1b shows the corresponding optical response under identical optical pumping conditions. In the unchained, randomly oriented configuration, the emission is dominated by broadband photoluminescence, indicating that the disordered medium does not provide sufficient scattering feedback to support lasing. As the nanowires are electrically assembled into aligned chains, sharp spectral peaks indicating the onset of random lasing begin to emerge. Furthermore, the lasing intensity depends sensitively on the orientation of the nanowire chains relative to the pump polarization: emission intensity is dramatically stronger when chains aligned parallel to the pump polarization. This behavior directly links the observed lasing response to the electrically defined scattering structure.

Taken together, these results establish a clear relationship between structure, scattering, and optical response. Electrical reconfiguration of nanowire assemblies enables tuning of structural disorder, which in turn controls the strength and efficiency of optical feedback in the random laser. Figure 1 therefore demonstrates that the optical response of a random lasing system can be engineered through active control over the inclusion structures, laying the foundation toward programmable random lasers based on reconfigurable scattering media.

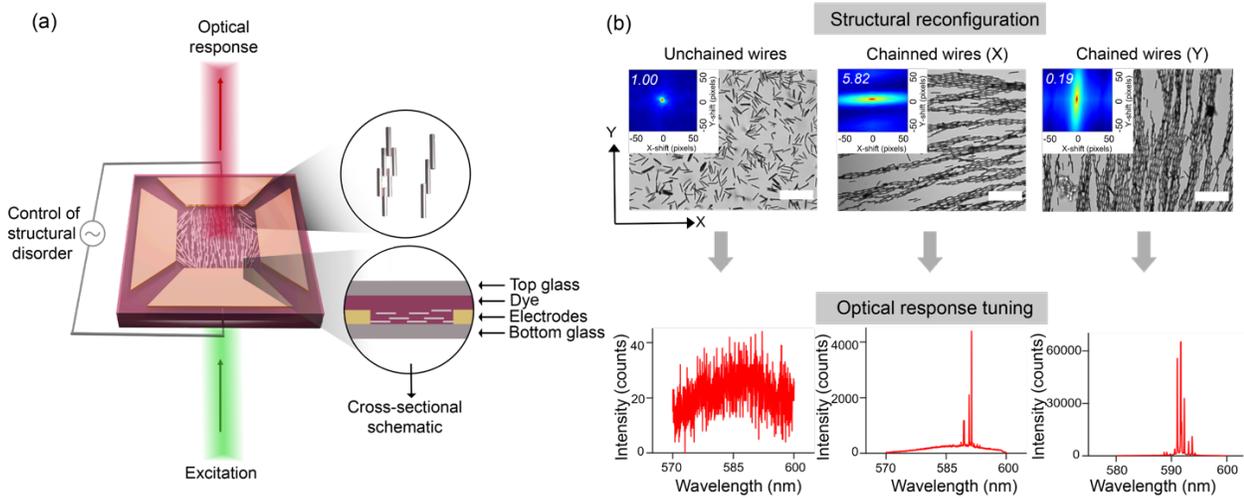

**Figure 1 | Electrically tunable structural disorder and its impact on random lasing.**
(a) Schematic illustration of the reconfigurable random lasing device. An external electric field is applied across patterned electrodes to actively control the structural configuration of metallic nanowires embedded in 2-mM rhodamine B dissolved in ethylene glycol. The device cross section shows the layered structure consisting of top glass, gain layer containing nanowires, electrodes, and bottom glass. Electrical actuation enables reversible tuning of the internal scattering architecture, while optical pumping excites random lasing emission. (b) Electrically controlled structural disorder and corresponding optical response. Top row: optical microscopy images of nanowire ensembles in three configurations, unchained and randomly oriented wires, nanowire chains aligned along the x direction, and nanowire chains aligned along the y direction. The white scale bar in each microscopy image corresponds to 20 μm. Insets show the corresponding two-dimensional autocorrelation maps, with numerical values quantifying the degree of orientational order. Bottom row: emission spectra recorded under identical excitation conditions (4.2 μJ pump energy). The unchained configuration exhibits only broadband photoluminescence without lasing signatures. In contrast, electrically assembled nanowire chains support random lasing, with emission intensities of approximately 4,000 counts for chains oriented perpendicular to the pump polarization and up to ~60,000 counts for chains aligned parallel to the pump polarization.

## 2.2 Reduced Random Lasing Threshold

Considering the strong structural anisotropy of the chained nanowire assemblies, we investigate the lasing effect when chaining orientation of the nanowires varies relative to the polarization of the linearly polarized pump light. **Figure 2** summarizes our experimental results, showing how electrically reconfigured nanowire assemblies modify the random-lasing threshold and emission characteristics. The top row of Figure 2(a-c) schematically illustrates the three structural configurations investigated: randomly oriented (no applied field), unchained nanowires; electrically chained nanowires aligned perpendicular to the pump polarization; and chained nanowires parallel to the pump polarization. The corresponding emission spectra recorded as a function of pump energy are shown in the bottom row of Figure 2(a-c).

For the unchained configuration (Figure 2a), the emission remains dominated by broadband photoluminescence over most of the excitation range. A weak lasing feature with intensity slightly exceeding 500 counts appears only at the highest pump energy of 7.2 µJ, indicating that the disordered medium provides insufficient optical feedback for low-threshold lasing. This behavior reflects the limited scattering strength and absence of collective feedback pathways in the unassembled nanowires. When an external electric field is applied to induce nanowire chaining perpendicular to the pump polarization (Figure 2b), the lasing behavior changes markedly. Under excitation energies between 0.6 and 4.2 µJ, sharp spectral peaks, a characteristic of random lasing, emerge, with peak intensities exceeding 4,000 counts at 4.2 µJ. The formation of nanowire chains increases the effective scattering cross section, leading to a substantial reduction in the lasing threshold compared with the unchained case. An even stronger effect is observed when the nanowire chains are aligned parallel to the pump polarization (Figure 2c). In this configuration, prominent lasing peaks with intensities nearly 10,000 counts appear at pump energies as low as 2.4 µJ. The pronounced enhancement indicates that the alignment parallel to the electric field of the incident light further boosts the efficiency of light-nanowire interaction, resulting in stronger optical feedback and the lowest lasing threshold among the three configurations.

The threshold analysis is summarized in Figure 2(d) and Figure 2(e). In Figure 2(d), the average intensity of the five strongest emission peaks is plotted as a function of pump energy, with a threshold criterion defined as an average peak intensity exceeding 500 counts. Using this metric, the lasing thresholds are determined to be 7.2 µJ for unchained nanowires, 4.2 µJ for chains oriented perpendicular to the pump polarization, and 2.4 µJ for chains oriented parallel to the pump polarization. Figure 2(e) provides confirmation based on spectral linewidth narrowing. Here, a full width at half maximum (FWHM) below 1 nm is used to identify the onset of lasing, yielding the same set of threshold energies for the three configurations. These results demonstrate that electrically induced nanowire chaining plays a dominant role in reducing the random-lasing threshold, and the relative orientation between the nanowire chains and the pump polarization provides an additional degree of control. These observations support the structure-determined optical response framework depicted in Figure 1 (see Materials and Methods, Data Collection and Processing).

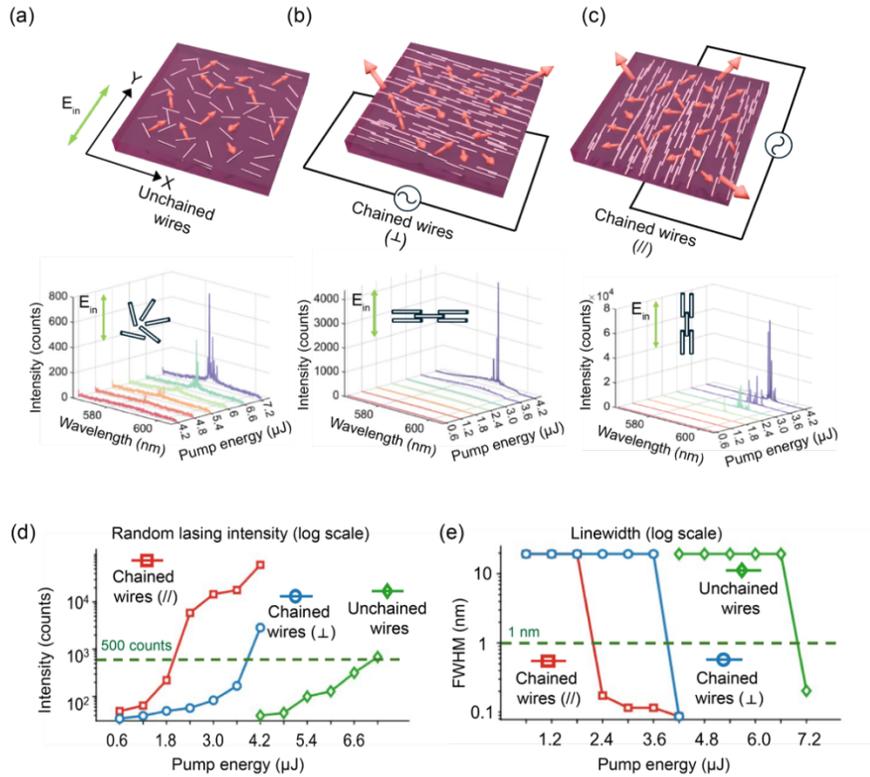

**Figure 2 | Random lasing performance governed by electrically reconfigured nanowire assemblies.**
(a-c) Top row: schematic illustrations of three nanowire configurations: (a) unchained and randomly oriented nanowires, (b) electrically chained nanowires aligned perpendicular (⊥) to the pump polarization, and (c) electrically chained nanowires aligned parallel (∥) to the pump polarization. The orientation of the incident pump electric field $E_{in}$ is indicated by the green arrow.
Bottom row: corresponding emission spectra recorded as a function of increasing pump energy for each configuration, showing the evolution from broadband photoluminescence to spectrally narrowed random lasing. (d) Average intensity of the five strongest emission peaks plotted as a function of pump energy. The random lasing threshold is defined as the pump energy at which the average peak intensity exceeds 500 counts (dashed line). Electrically induced nanowire chaining leads to a substantial reduction in lasing threshold compared with the unchained case. (e) Full width at half maximum (FWHM) of the strongest emission peak as a function of pump energy. A linewidth below 1 nm (dashed line) is used as the criterion for lasing onset. The threshold values extracted from the FWHM analysis are consistent with those obtained from the intensity criterion.

## 2.3 Assembly Induced Enhancement of Scattering

In our random laser, photon confinement is provided by scattering from the nanowire assembly, whereas competing loss channels include absorption by the nanowires and other dissipative processes. A useful metric for comparing assembly configurations is the ratio between the scattering cross-section and the total extinction cross-section (S/T ratio), $\sigma_s / (\sigma_s + \sigma_a)$. **Figure 3** presents the simulation results for the three configurations introduced in Figure 2, modeled in idealized form to capture the key geometric features of the experimentally observed assemblies (see Materials and Methods, Numerical Simulation).

Figure 3a shows the simulated scattering efficiency as a function of nanowire overlap length (OL). Within a chain, adjacent nanowires may partially overlap along their long axes, and we define OL as the axial overlap between neighboring wires. Larger OL indicates progressively stronger chaining. The S/T ratio shows clear polarization dependence across all OL values: the ratio of the parallel case is significantly larger than that of the perpendicular case. In addition, the S/T ratio increases as OL transitions from negative (separated wires) to positive (overlapped wires) values, likely due to enhanced electromagnetic coupling between neighboring wires. Figure 3b further examines the S/T ratio across the wavelength range of interest, spanning both the pump wavelength and the emission band of the gain medium. Here, 532 nm corresponds to the pump wavelength, while 585 nm is near the central peak wavelength of the dye emission. At a representative overlap length of OL = 1 μm, chains aligned parallel (∥) to the incident electric-field polarization exhibit substantially higher scattering efficiency than chains aligned perpendicular (⊥) throughout this spectral region. The enhanced scattering efficiency at 532 nm indicates more effective pump-light trapping and energy deposition within the gain region, while the enhancement near 585 nm implies stronger radiative feedback for stimulated emission. Together, this wavelength- and polarization-dependent scattering efficiency analysis reveals the underlying mechanism of the anisotropic lasing behavior observed experimentally (see Materials and Methods, Numerical Simulation).

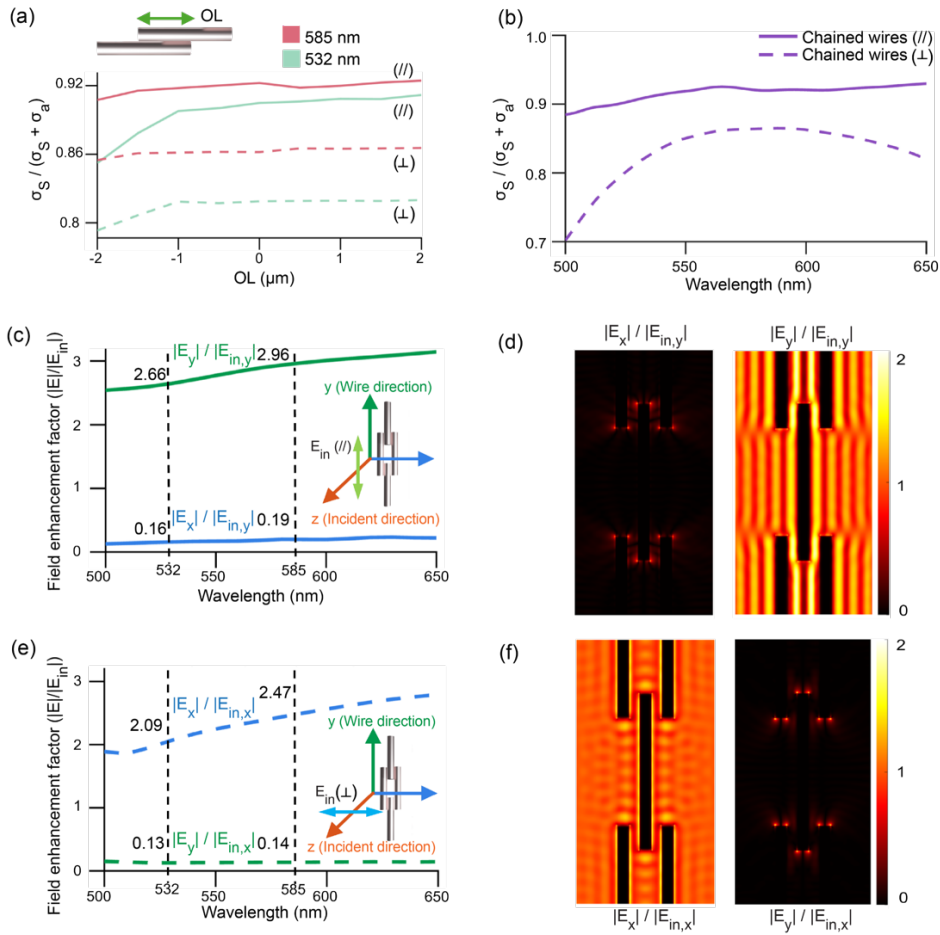

**Figure 3 | Scattering enhancement and polarization-dependent near-field response induced by nanowire chaining (numerical simulations).** (a) S/T ratio versus nanowire overlap length (OL). Calculated ratio of the scattering coefficient to the total extinction coefficient (scattering + absorption) $\sigma_s / (\sigma_s+\sigma_a)$, as a function of nanowire overlap length. Larger OL indicates increasing degrees of nanowire

chaining. The increasing trend demonstrates that nanowire assembly enhances radiative scattering relative to absorption as electromagnetic coupling between wires increases. (b) Scattering efficiency at fixed overlap length. Comparison at a fixed overlap length of OL = 1 µm, showing that nanowire chains aligned parallel (∥) to the incident electric-field polarization exhibit a higher S/T ratio, than chains aligned perpendicular (⊥). This behavior indicates reduced dissipative loss and more efficient radiative feedback for parallel alignment. (c) Near-field enhancement for nanowire chains aligned parallel to the incident electric field. Spectrally resolved near-field enhancement factors $|E_y| / |E_{in,y}|$ and $|E_x| / |E_{in,y}|$ evaluated at OL = 1 µm. The electric-field component parallel to the nanowire axis dominates the local response, indicating strong polarization-selective near-field enhancement. (d) Corresponding near-field intensity distributions for parallel alignment. Simulated near-field intensity maps at the emission wavelength (585 nm), showing pronounced field enhancement distributed along the nanowire chains, consistent with efficient coupling between the optical field and the chained nanowire structure. (e) Near-field enhancement for nanowire chains aligned perpendicular to the incident electric field. Spectrally resolved near-field enhancement factors $|E_x| / |E_{in,x}|$ and $|E_y| / |E_{in,x}|$ at OL = 1 µm. Compared with the parallel configuration, the overall enhancement is reduced, indicating weaker coupling. (f) Corresponding near-field intensity distributions for perpendicular alignment. Simulated near-field intensity maps at 585 nm, revealing weaker field enhancement and less efficient interaction with the nanowire structure relative to the parallel case.

**2.4 Assembly Induced Enhancement of Near Field Effects**

Nanowire chaining also reshapes the local electromagnetic environment experienced by both the pump field and the lasing field in the gain medium. Figure 3(c-f) summarizes the evaluations of the simulated the near-field results. The spectra of the enhancement factors ($|E_x|/|E_{in}|$ and $|E_y|/|E_{in}|$ where $E_x$ and $E_y$ are the x- and y-components of the electric field and $|E_{in}|$ is magnitude of the incident wave) when polarization of the incident light is parallel (perpendicular) to the chaining direction are shown in Fig. 3(c) (Fig.3-e), and the corresponding field distributions at 585 nm are shown in Fig. 3(d) (Fig. 3-f).

For nanowire chains aligned parallel to the incident electric field, Figure 3(c) shows a strongly anisotropic response in which the field component along the nanowire axis dominates. In particular, at 532 nm (585 nm), the axial enhancement reaches 2.66 (2.96), whereas the orthogonal component remains much smaller, i.e., 0.16 (0.19). The large contrast between parallel and perpendicular field components indicates efficient electromagnetic coupling enabled by the aligned nanowire chain geometry, resulting in substantial local field component along the chaining direction. Consistent with this behavior, the corresponding near-field intensity map at 585 nm illustrated in Figure 3(d) shows pronounced field enhancement effect in the vicinity of the chained structure. When the nanowire chains are aligned perpendicular to the incident electric field, the near-field enhancement is reduced at both wavelengths. As shown in Figure 3(e), the dominant enhancement factor decreases to 2.09 at 532 nm and 2.47 at 585 nm, while the orthogonal component remains weak (0.13 at 532 nm and 0.14 at 585 nm). Compared with the parallel case, both the overall enhancement and the polarization selectivity are reduced. The near-field intensity map at 585 nm in Figure 3(f) further confirms this trend, showing weaker and less spatially extended field enhancement (See Supporting Information, Figure S3 for Near-filed Intensity Map at 532nm). Figure 3(c-f) demonstrates that nanowire chaining produces an orientation-dependent near-field response that affects both the pump absorption at 532 nm and the lasing emission near 585 nm wavelength. Compared with the perpendicular alignment, parallel alignment consistently yields stronger local field enhancement, supporting the enhanced optical feedback and improved random lasing performance observed in the experiments (See Materials and Methods, Numerical Simulation).

## 2.5 Control Over Polarization of Random Lasing

Our platform further enables electrical control over the polarization state of the random-lasing emission. To isolate the role of nanowire orientation, circularly polarized light was used to pump the system, in which the polarization properties of the emission are primarily governed by the internal scattering structure. A schematic of the experimental setup is illustrated in **Figure 4a**, where the anisotropic gain medium due to the chaining nanowires is pumped by circularly polarized light. The random lasing emission develops a pronounced polarization preference that reflects the nanowire chaining direction. To quantify the effect, we used a polarization-sensitive camera to simultaneously record the *x*- and *y*-polarized components of the emitted random lasing for the three nanowire configurations (**Figure 4b**). For the unchained scenario, randomly oriented nanowires were excited at an excitation level of 7.2 µJ. The corresponding emission intensities in the two polarization channels are nearly identical, yielding a polarization ratio of 1.01. This indicates that the lasing emission exhibits no measurable polarization preference when the scattering structure is isotropic. In contrast, when the nanowires are electrically assembled into chains aligned along the *x* direction and excited at 4.2 µJ, the *x*-polarized emission component becomes significantly stronger than the *y*-polarized component, resulting in a polarization ratio of 2.66. Similarly, for nanowire chains aligned along the *y* direction, the *y*-polarized emission dominates, with a polarization ratio of 1.56. Note that residual polarization-dependent transmission within the microscope optics, which is not readily accessible for adjustment, may also have contributed a small systematic offset to the measured polarization ratios. (See Materials and Methods: Data Collection and Processing).

The corresponding pixel-intensity histograms in **Figure 4c** further confirm this behavior. For the unchained configuration, the intensity distributions of the two polarization channels largely overlap. In contrast, for the electrically assembled samples, the distribution corresponding to the polarization aligned with the nanowire chains extends to higher intensities, indicating a preference along that direction. The polarization ratios are extracted by comparing the integrated intensities of the top 10% brightest pixels in each polarization channel, providing a robust metric that is not sensitive to background fluctuations. These results establish that electrically assembled nanowire chains impart a structural anisotropy that manifests as a polarization preference in random lasing emission. This behavior directly links the polarization characteristics of the random laser to the internal scattering geometry, providing an additional degree of freedom for tailoring emission properties through active structural control.

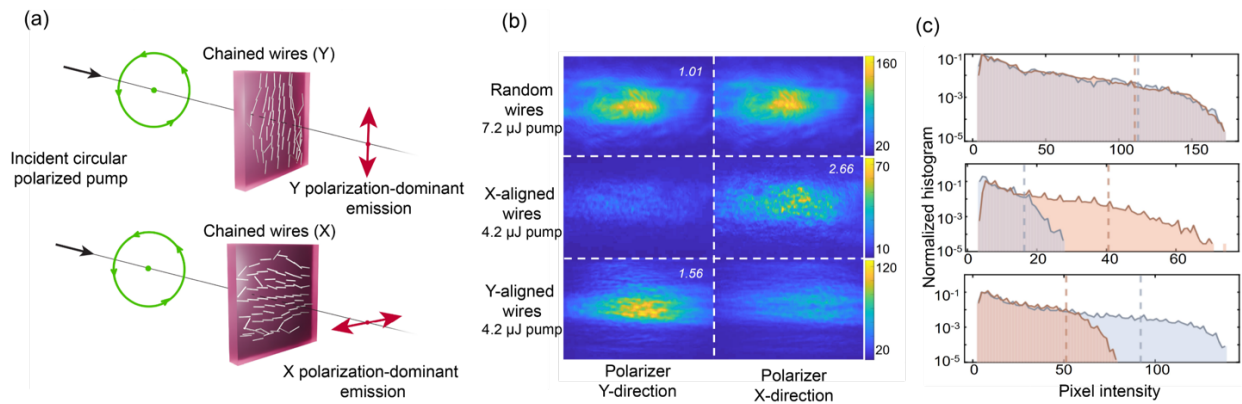

**Figure 4 | Polarization preference in random lasing enabled by electrically aligned nanowire assemblies under circularly polarized excitation.**
(a) Conceptual illustration of polarization-selective random lasing enabled by electrically assembled nanowire chains. The gain medium is excited by a circularly polarized pump, ensuring equal excitation of orthogonal field components. The emitted random lasing exhibits a pronounced polarization preference aligned with the nanowire chaining direction. Two representative configurations are shown for nanowire chains aligned along the x and y directions, respectively, indicating that the emission polarization is governed by the internal scattering anisotropy.
(b) Polarization-resolved images acquired with a polarization-sensitive camera, showing the spatial distributions of the y-polarized (left column) and x-polarized (right column) components of the random lasing emission. Three nanowire configurations are shown: unchained and randomly oriented nanowires excited with a 7.2 µJ pump (top row), nanowire chains aligned along the x direction excited with a 4.2 µJ pump (middle row), and nanowire chains aligned along the y direction excited with a 4.2 µJ pump (bottom row). The polarization ratio, calculated from the top 10% brightest pixels in each polarization channel, is annotated for each configuration (1.01, 2.66, and 1.56, respectively).
(c) Log-scaled histograms of normalized pixel intensities corresponding to the polarization-resolved images in (b). Orange and blue curves represent the x- and y-polarized emission components, respectively. Vertical dashed lines indicate the intensity thresholds used to define the top 10% brightest pixels for polarization-ratio calculation.

## 3. Conclusion

Random lasing provides a promising route toward compact, low-cost, and cavity-free light sources with potential applications in wearable devices, biosensors, and lab-on-chip technologies. In this work, we have demonstrated an electrically reconfigurable random-lasing platform with actively programmed scattering topology. By applying an AC electric field across quadrupole electrodes, we induced dielectrophoretic chaining and alignment of Ag nanowires dispersed in a dye gain medium, enabling reconfiguration of the internal disorder. This structural control allowed simultaneous tuning of key lasing characteristics, including threshold behavior, emission intensity, and polarization. Experiments showed that nanowire chaining and alignment enhance scattering-mediated feedback and reduce the lasing threshold, with the lowest threshold observed when the chain direction is aligned parallel to the pump polarization. Consistent with this trend, simulations indicate the increased scattering efficiency for stronger chaining with polarization dependent field enhancement effect, providing a mechanistic basis for the experimental observations. Finally, using circularly polarized pumping we have demonstrated that nanowire orientation induces a clear polarization preference in the random-lasing output, establishing a practical route to polarization-selective random lasers. Together, these results highlight electrically programmable disorder as a versatile strategy for actively tuning random lasing and enabling reconfigurable photonic functionalities.

## 4. Materials and Methods

### 4.1 Electrode Fabrication

Gold quadrupole electrodes with a 200 µm gap were fabricated on No. 2 glass coverslips (35 mm diameter, GlycoTech) via photolithography as follows.[48–50] A bi-layer photoresist (LOR 5A and SPR-3012, MicroChem) was spin-coated onto the coverslips in which each spin coat was followed by baking at 200 °C for 3 minutes and baking at 95 °C for 1 minute, respectively. Then the coverslips were patterned with the desired Quadruple design using a Heidelberg MLA 150 direct-write system. The patterned coverslips were developed in MF-CD-26 developer (2.4% tetramethylammonium hydroxide, MicroChem) for 75 s followed by washing with DI water and drying with $N_2(g)$. Subsequently, a 15 nm titanium adhesion layer followed by 60 nm gold was deposited using a Temescal FC2000 evaporation system. For the Lift-off, first, the coverslips were immersed in acetone for 30 minutes to ensure effective dissolving of the resist. Next, the coverslips were placed in a holder and immersed in a glass container filled with Baker PRS 3000 photoresist stripper (MicroChem). The container was then immersed in a hot water bath at 80°C for an hour until the lift off was complete. Finally, the coverslips were washed with DI water and dried with $N_2(g)$.

### 4.2 Silver Nanowire Synthesis

Silver nanowires were synthesized using a previously described templated electrodeposition method.[51] First, 0.2 μm pore diameter Whatman alumina Anodisc filter membranes were evaporated with Ag on the branched side of the membrane, using a Temescal FC2000 evaporation system to a thickness of 300 nm. This makes electrical contact between the membrane and the Ag electrode used for electrodeposition. Next, a commercially available Silver plating solution (Technic 1025 RTU) was electroplated into 0.2 μm pore diameter Whatman alumina Anodisc filter membranes at a current density of 1.01 mA/cm$^2$ to ensure that the pores inside are also filled and no leaks will occur from the back electrode. For the nanowire deposition, two gold tips were deposited (at a current density of 0.33 mA/cm$^2$) at the two ends with the Ag nanowire in the middle.

Au tips were used to protect the Silver wire from etching in the next step.[52,53] The pore size and deposition time determined the nanowire diameter and length, respectively. After the desired deposition time, the silver electrode backing on the membrane was etched by briefly immersing the membrane in 33 % nitric Acid. Then the membranes were washed with DI water and transferred to a centrifuge tube containing 3 M NaOH and vortexed for 25 minutes to release the wires from the membrane. After removing the membrane outer plastic ring, the suspension was centrifuged to separate the nanowires. The nanowires were washed with water twice, followed by two ethanol washings while centrifuging between each washing. Finally, the nanowires were coated twice to reach an approximate 200 nm layer of amorphous silica via a previously reported sol–gel process with a small modification where we used NaOH instead of Ammonium hydroxide to avoid Ag nanowire etching during the glass-coating process.[54] Schematic demonstration is provided in the Supplemental Information (See Supporting Information, Figure S4).

### 4.3 Directed Self Assembly of Silica-coated Ag Nanowires

The silica-coated silver nanowires were then dispersed in a 2 mM rhodamine B solution in ethylene glycol at a concentration of $1 \times 10^8$ wires/mL. Wire concentration was determined by counting the wires in a given volume using a hemocytometer. Random laser sample containing the silver wires dispersed in the dye medium was mounted on a microscope stage. For quadrupole electrode experiments, a 500 μm thick silicone spacer was affixed to the electrode such that the hole in the spacer was centered over the electrode gap, forming a well to contain ~3.5 μL of the nanowire solution. An index-matching oil droplet (Cargille Immersion Oil, type HF) was applied to the coverslip, with an additional spacer placed on a top coverslip to prevent cavity lasing.

Electrical connections to each of the four quadrupole electrode pads were established using gold thread (Ametek Electronic Components) and silver adhesive (Electron Microscopy Sciences). Two pads were connected to each of two separate function generators, enabling independent control of the applied field along the x and y axes. Activating the generator connected to a given axis induced nanowire chaining along that axis.

The applied field conditions and nanowire concentration ($1 \times 10^8$ wires/mL) were selected to promote chain formation across the electrode gap. Alternating between "on" and "off" states of the field enabled reversible assembly and disassembly of the nanowire chains. For nanowires in rhodamine B/ethylene glycol solution, AC field parameters of 3 MHz and 1250 V/cm produced the most uniform chaining and alignment.

### 4.4 Experimental Optical System

To perform reconfigurable random lasing experiments, we designed and built a custom optical system by integrating a 532 nm, 5 ns, 10 Hz Nd:YAG pulsed laser (Quantel) with a Nikon TE2000-U inverted microscope and a high-resolution spectrometer (SpectraPro 2500i, Princeton Instruments) coupled to a liquid-nitrogen-cooled CCD camera (See Supporting Information, Figure S5). The laser power was controlled using a half-wave plate and a polarizing beam splitter. The half-wave plate adjusts the polarization of the incident laser beam, thereby altering the power ratio between the s- and p-polarized components. Since the polarizing beam splitter transmits only the p-polarized component and reflects the s-polarized light, this configuration enables control of the excitation pulse energy entering the system. To control the excitation spot size, the laser beam was expanded using a lens pair consisting of a convex lens with a 150 mm focal length and a concave lens with a 100 mm focal length. The expanded beam was then coupled into the microscope's side port and focused onto the sample plane using a 100× oil-immersion objective lens, resulting in an excitation spot diameter of approximately 60 μm. Emission from the sample was collected through the same objective lens and directed to two parallel detection paths. A 50:50 non-polarizing beam splitter guided the emission to both a polarization-sensitive camera and a spectrometer. A reflective diffraction grating was used to spatially separate the excitation laser from the emission signal. To ensure that only emission was detected, a 550 nm long-pass filter was placed in front of the polarization camera. In the spectrometer path, the emission was coupled into a multimode optical fiber. Before entering the fiber, the emission passed through another 550 nm long-pass filter to block residual pump light and ensure clean spectral acquisition.

### 4.5 Numerical Simulation

Full-wave electromagnetic simulations were performed using CST Microwave Studio, a commercial finite integration package. The modeled geometry in all cases was an isolated nanowire chain surrounded by a solution (Ethylene Glycol, with a refractive index of 1.45) domain. To mimic the infinitely long nanowire chain, periodic boundary condition was applied to vertical facets of the simulation domain. To obtain the scattering cross-section of the structures, the normal scattered Poynting vector was integrated on a closed surface that encloses an individual nanowire chain illuminated by a broadband plane wave. To quantify the field enhancement effect, near-field evaluations were performed on the outer surfaces of the nanowires. To correlate the model to experimental conditions, widely accepted experimental material properties of silver and gold were adapted for all simulations. Additionally, the refractive index of silica is assumed to be 1.45.

### 4.6 Data Collection and Processing

The wire chaining autocorrelation results were obtained by first capturing alignment images of the nanowire assemblies using an optical microscope. Each grayscale image was binarized using a 50% intensity threshold to isolate the nanowire structures from the background. Two-dimensional autocorrelation was then computed using MATLAB in both the x and y directions. The resulting autocorrelation map contains a central peak representing the maximum correlation, corresponding to the case where the image is perfectly overlapped with itself. To quantify alignment anisotropy, we defined the correlation radius as the distance from the center at which the autocorrelation intensity drops by 30% along both the x and y axes. This radius reflects the shift, in pixels, required to reduce the structural similarity between the original and shifted images. The ratio of correlation radii in the x and y directions was used as a quantitative metric to assess the degree of nanowire alignment. A ratio near 1 indicates isotropic distribution, while higher or lower values reflect strong anisotropic alignment along specific directions.

The cover glass and bottom substrate together may form a Fabry-Pérot cavity, potentially introducing cavity-enhanced lasing modes. To suppress this effect and ensure that only random lasing is measured, a droplet of index-matching oil is applied to the top surface of the cover glass before each measurement. The dome-shaped oil droplet minimizes the refractive index contrast at the top interface, effectively suppressing the Fabry-Pérot effect. When the incident pump energy exceeds the random lasing threshold, stimulated emission is generated within the scattering medium. The excitation source operates in single-shot mode, ensuring that each measurement captures the lasing response to an individual pump pulse. After oil application, we validate that Fabry-Pérot cavity lasing cannot be excited under the given pump conditions. Furthermore, the random lasing spectrum is monitored using the spectrometer to confirm the absence of periodic spectral features, ensuring that the observed emission exhibits stochastic peak distributions characteristic of random lasing. To minimize the possibility of dye bleaching, the excitation location on the sample was frequently changed during the measurements.

To ensure accurate measurement of the lasing threshold, a lasing event is considered effective only if the observed peak intensity exceeds the photoluminescence background by at least 500 counts. If no effective random lasing is detected at a given pump energy, the measurement is repeated three times at different locations on the sample using the same pump energy. This repetition helps confirm that the energy is indeed below the lasing threshold and rules out the possibility that a localized sample variation might prevent lasing even when the excitation is sufficient. Although the exact lasing threshold values may vary slightly, due to factors such as pump energy fluctuations, local sample inhomogeneity, power meter uncertainty, or human error, a consistent trend is observed across all conditions. Samples with unchained nanowires and random orientation exhibit the highest lasing threshold. The formation of nanowire chains lowers the threshold, and further reduction is achieved when the chaining direction is aligned parallel to the linear polarization of the pump. This trend is attributed to enhanced light-matter interaction and an increase in the effective scattering cross-sectional area.

To process the polarization camera data, we first isolate the random lasing signal from background contributions such as photoluminescence and noise. This is done by identifying which polarization channel, x or y, has the higher total intensity across all pixels in the image. From that channel (e.g., the y-polarized image), we then select the top 10% of pixels with the highest intensity values. These high-intensity pixels are assumed to be dominated by random lasing emission. The same pixel locations are then used to extract corresponding intensity values from the other polarization image (e.g., the x-polarized image), ensuring a consistent basis for comparison. The polarization ratio is calculated by dividing the total intensity of the selected top 10% pixels in one polarization image by the total intensity from the same pixel locations in the orthogonal polarization image.

## Acknowledgments

This work was funded by the National Science Foundation ECCS- 2303189. Partial support for this work was also provided by the John L. and Genevieve H. McCain endowed chair professorship at the Pennsylvania State University.

## Author Contributions

*Jinkai Yang* and *Kumudu N. Ranasinghe* contributed equally to this work.

## Data Availability Statement

Data will be made available upon reasonable request.

## Conflicts of Interest

The authors declare no conflicts of interest.

# Supporting Information

**Taming Randomness in Random Lasers: Programmable Disorder for Active Control of Random Lasing via Electric-Field-Directed Assembly of Nanowires**

**Authors**

*Jinkai Yang,[1,2] Kumudu N. Ranasinghe,[3] Lei Kang,[1,\*] Jennifer R. Decker,[3] Cheng-Yu Wang,[1,2] Douglas H. Werner,[1] Christine D. Keating[3,\*] and Zhiwen Liu[1,2,\*]*

**Affiliations**

1. School of Electrical Engineering and Computer Science, The Pennsylvania State University, University Park, Pennsylvania 16802, United States.
2. Materials Research Institute, The Pennsylvania State University, University Park, Pennsylvania 16802, United States.
3. Department of Chemistry, The Pennsylvania State University, University Park, Pennsylvania 16802, United States.

**Email Address**

Jinkai Yang: jxy208@psu.edu
Kumudu N. Ranasinghe: knr5313@psu.edu
Lei Kang: lzk12@psu.edu
Jennifer R. Decker: jrdecker01@ysu.edu
Cheng-Yu Wang: cuw70@psu.edu
Douglas H. Werner: dhw@psu.edu
Christine D. Keating: keating@chem.psu.edu
Zhiwen Liu: zzl1@psu.edu

**\* Corresponding authors**
Lei Kang: lzk12@psu.edu
Christine D. Keating: keating@chem.psu.edu
Zhiwen Liu: zzl1@psu.edu

**Funding**

This work was funded by the National Science Foundation ECCS- 2303189. Partial support for this work was also provided by the John L. and Genevieve H. McCain endowed chair professorship at The Pennsylvania State University.

**Author Contributions**

*Jinkai Yang* and *Kumudu N. Ranasinghe* contributed equally to this work.

**Conflicts of Interest**

The authors declare no conflicts of interest.

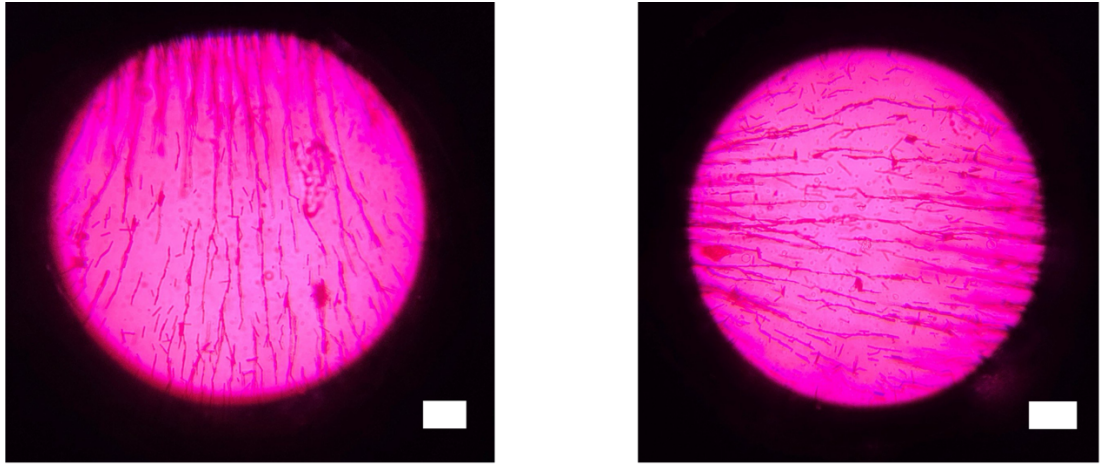

Figure S1. Images of Ag nanowires dispersed in a 2 mM rhodamine B/ethylene glycol (EG) solution, forming chain-like structures under an applied electric field of 3 MHz and 1250 V/cm. The nanowire chains can be assembled along both the X and Y directions as the direction of the applied electric field is changed.(Scale bar = 20 µm)

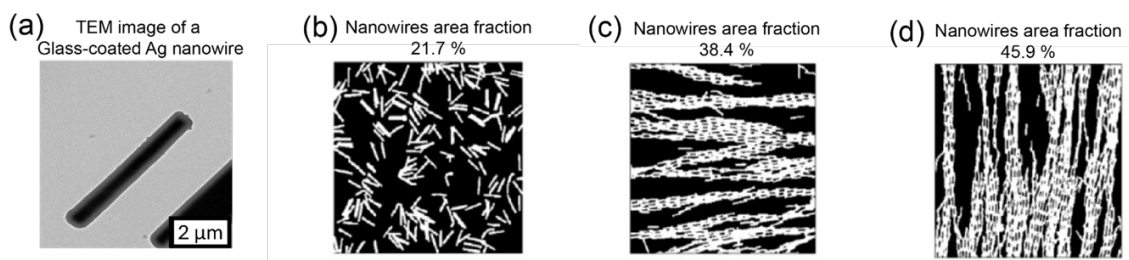

Figure S2. (a) Transmission electron microscopy (TEM) image of a representative glass-coated silver nanowire used in this study, showing uniform morphology and smooth coating. Scale bar: 2 µm. (b–d) Binary-thresholded optical microscopy images used to quantify nanowire area fraction under different assembly conditions. (b) Unchained, randomly oriented nanowires exhibit a low area fraction of 21.7%. (c) Electrically assembled nanowire chains show an increased area fraction of 38.4%. (d) Stronger electrical assembly further increases the local nanowire area fraction to 45.9%. The progressive increase in area fraction reflects electrically induced nanowire densification within the electrode region, which enhances the effective scattering strength of the medium.

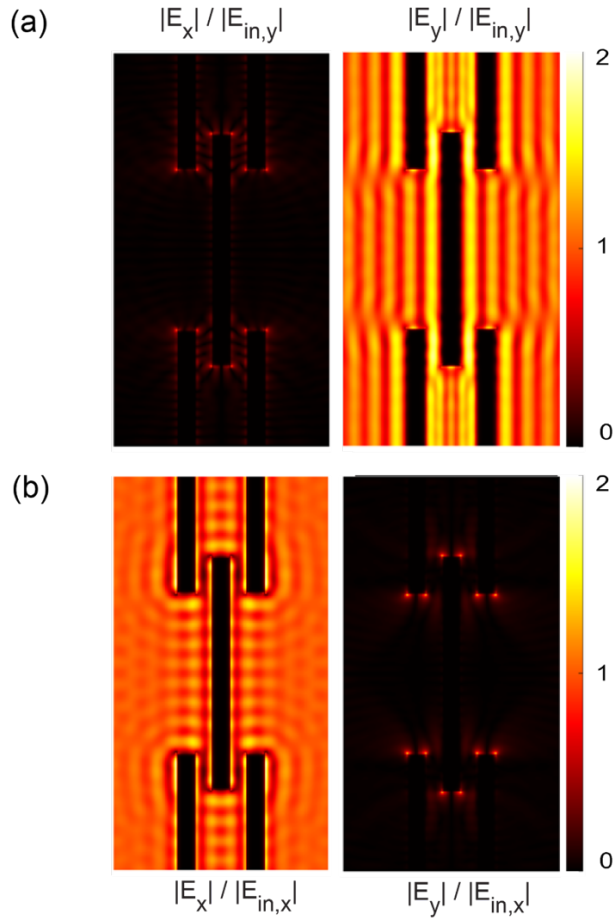

Figure S3. (a) Simulated near-field intensity distributions for nanowire chains aligned parallel to the incident electric-field polarization at 532 nm. The dominant field component is oriented along the nanowire axis, resulting in stronger local field enhancement concentrated around the chained nanowire structure. (b) Simulated near-field intensity distributions for nanowire chains aligned perpendicular to the incident electric-field polarization at 532 nm. In this configuration, the overall near-field enhancement is weaker and more spatially confined, indicating reduced electromagnetic coupling between the incident field and the nanowire chains. These simulations at the pump wavelength complement the 585 nm near-field results shown in Fig. 2(d,f) and confirm that polarization-dependent nanowire alignment influences the local electromagnetic response at both excitation and emission wavelengths.

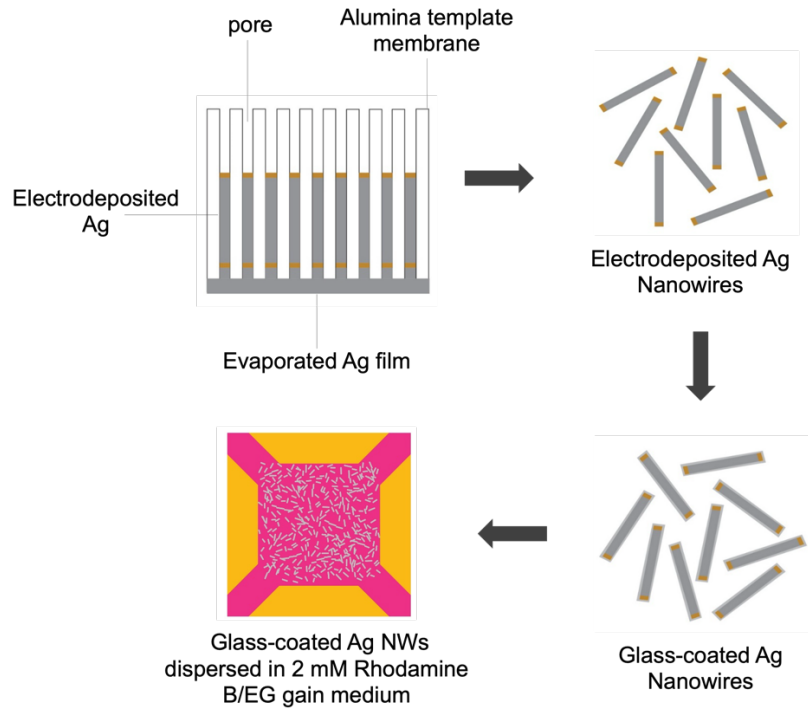

Figure S4. Schematic illustrating the synthesis of Ag nanowires using a templated electrodeposition method. The nanowires are subsequently glass-coated, dispersed in a 2 mM rhodamine B/ethylene glycol solution, and transferred to the electrode setup, where they can be manipulated under an applied electric field.

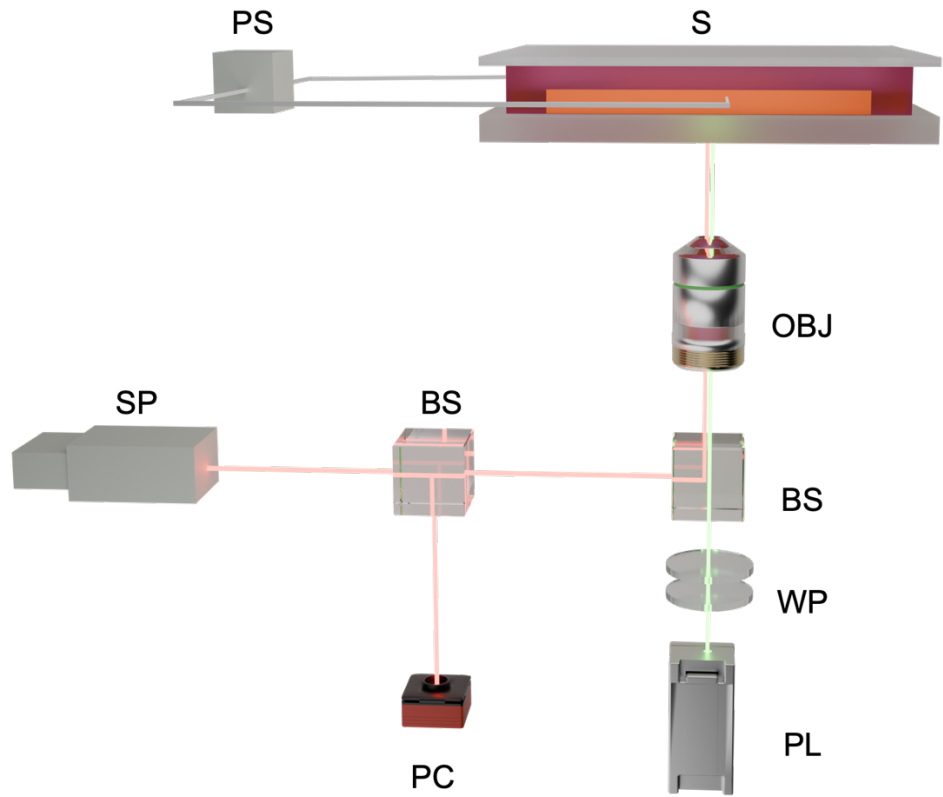

Figure S5. Schematic of the experimental optical setup used for excitation and emission characterization. The pump laser (PL) passes through wave plates (WP) to control the excitation polarization and is directed toward the sample (S) through a beam splitter (BS) and objective lens (OBJ). The emitted light is collected by the same objective and separated by beam splitters for simultaneous detection by a polarization camera (PC) and a spectrometer (SP). A power supply (PS) applies an external electric field across the sample electrodes to enable electrical reconfiguration of the nanowire assemblies during optical measurements. Optical components used in the Blender visualizations were sourced from the asset set by Ryo Mizuta Graphics.